\documentclass{ws-mpla}

\usepackage{graphicx,epsfig}

\usepackage{xcolor}
\usepackage{topcapt}
\newcommand{\ci}[1]{\cite{#1}}
\newcommand{\lab}[1]{\label{#1}}

\newcommand{\ba}{\begin{eqnarray}}
\newcommand{\ea}{\end{eqnarray}}
\newcommand{\beqs}{\begin{eqnarray}}
\newcommand{\eeqs}{\end{eqnarray}}
\newcommand{\lsim}{\mathrel{\raisebox{-.6ex}{ $\stackrel{\textstyle<}{\sim} $}}}
\newcommand{\gsim}{\mathrel{\raisebox{-.6ex}{ $\stackrel{\textstyle>}{\sim} $}}}

\newcommand{\dd}{\Delta}
\newcommand{\psim}{{{}<\atop{}^\sim}}
\newcommand{\ua}{\sigma(^{\uparrow})}
\newcommand{\uu}{\sigma(^{\uparrow \uparrow})}
\newcommand{\da}{\sigma(^{\downarrow})    }
\newcommand{\ud}{\sigma(^{\uparrow \downarrow})  }
\newcommand{\dds}{\sigma(^{\downarrow \downarrow}) }
\newcommand{\du}{\sigma(^{\downarrow \uparrow})    }
\newcommand{\dss}{\Delta \sigma^{s} }
\newcommand{\dws}{\Delta \sigma^{d}}
\newcommand{\sq}{\sqrt{s}}
\newcommand{\stot}{\sigma_{tot}(s)}

\begin{document}


%
\title{Anomaly in the differential cross sections at 13 TeV
}

 \author{
O. V. Selyugin\thanks{\email{selugin@theor.jinr.ru}  }
}
%
\address{
 BLTP,
Joint Institute for Nuclear Research,
141980 Dubna, Moscow region, Russia }
%


\maketitle


\begin{abstract}
The analysis of the new TOTEM data at 13 TeV in a wide momentum transfer region
reveals the unusual phenomenon - the presence in the elastic scattering amplitude
of a  term with a very large slope that is responsible for the behavior of  hadron
scattering at a very small momentum transfer. This term can be connected with  hadron
interactions at large distances.
\end{abstract}

\ccode{
      13.40.Gp, 14.20.Dh, 12.38.Lg }



%

 \section{Introduction}

      The new data of
     the  TOTEM Collaboration on the elastic differential cross sections
   at 13 TeV have two sets of  data - at small momentum transfer \cite{T66}
   and at middle and large momentum transfer \cite{T67}.
   Recently, the first set of data has created  a wide discussion of the determination of the total
   cross section and the value of $\rho(t=0)$   (for example \cite{MN,Khoze,CS-Diff}).
    A  research of the structure of the elastic hadron scattering amplitude
   at superhigh energies and small momentum transfer - $t$
  can give a connection   between
  the  experimental knowledge and  the basic asymptotic theorems
    based on  first principles \cite{akm,fp,royt}.
   It gives   information about the  hadron interaction
   at large distances where the perturbative QCD does not work \cite{Drem},
   and a new theory as, for example, instanton or string theories
    must be developed.

           Usually,  a small region of  $t$ is taken into account
   for extraction of the sizes of $\sigma_{tot}$ and $\rho(t=0)$ (for example \cite{T66,Protvino19}).
   Really, already in the analysis of the UA4/2 data it was shown that the value of $\rho(s,t)$
   has a phenomenological meaning, as its determination requires some model assumptions \cite{Sel-UA42}.
   A simple exponential approximation of the data gave $\rho=0.24$ from the UA4 data and $\rho=0.129$
   from UA4/2 data (both at $\sqrt{s} =540$ GeV. More complicated analyses gave
   $\rho=0.19$ from the UA4 data and $\rho=0.16$ from UA4/2 data \cite{Sel-UA42}.
   Hence, this is  not an experimental problem
   but a theoretical one  \cite{CS-PRL09}.
   A phenomenological form of the scattering amplitude  determined for small $t$ can lead
   to  very different differential cross sections at larger $t$.
   Especially, it is connected with the differential cross section at 13 TeV, as the diffraction minimum
   is located  at a non-large $t$.

   Also, a very important moment is related with the question how the experimental uncertainty,
   which is usually  named  experimental errors, is used in our fitting procedure.
    In fact, the actual background rates and shapes of the measured distributions are sensitive
 to a number of experimental quantities such as calibration constants, detector geometries,
  poorly known material budgets within experiments, particle identification efficiencies,  etc.
   A 'systematic error', referred to by a high energy physicist, usually corresponds
    to a 'nuisance parameter' by a statistician.

   Hence, the extraction of the main value of the elastic hadron interaction
   requires some model that can describe all experimental data
   at the quantitative  level with minimum free parameters.
   Now  many groups of researchers have presented some physical models satisfying
   more or less  these   requirements. It is  especially   related with the HEGS
   (High Energy Generalized Structure) model \cite{HEGS0,HEGS1}.
   As it takes into account  two form factors (electromagnetic and gravitomagnetic),
   which are calculated from the GPDs function of nucleons, it has a minimum free
   parameters and gives a quantitative description of the exiting experimental data
   in a wide energy region and momentum transfer.
   Analysis of new data  of the TOTEM Collaboration at 13 TeV in the framework of the HEGS model
     discovered a new
   phenomenon in the hadron interaction - the oscillation term of the elastic scattering
   amplitude \cite{Osc-13}. During this analysis only statistical errors of  experimental
   data were taken into account in the fitting procedure. Systematic errors were taken as
    an additional coefficient of the normalization of the differential cross section,
   which is  independent of the momentum transfer.

   Further careful analysis of the behavior of the differential cross sections
   in the framework of the HEGS model have shown additional unusual properties of the
   behavior of the elastic scattering amplitude at a very small momentum transfer.
   The effect is  examined  from different  points of view in the present
   paper.

   In the second section of the paper, the new effect is analyzed in the framework of the HEGS model
   with taking into account  experimental data of both the sets of the TOTEM Collaboration
   obtained at 13 TeV and is compared with the results of some other models in the third section.
   In the fourth section, the existence of the new effect is examined in a simple phenomenological
   form of the scattering amplitude (as used most groups of researchers) and
   experimental data of only the first set at small momentum transfer  are taken  into account.
   The conclusions are given in the final section.

\section{Some problems in the description of the differential cross section
 in a wide region of momentum transfer}

    There are many different semi-phenomenological models which give a qualitative
    description of the behavior of the differential cross sections of the elastic proton-proton
    scattering at $\sqrt{s} = 13$ TeV (for example \cite{Kohara,Gotsman,Koh-Bl18}).
    Some examples can be found in the review \cite{Pakanoni}; hence, we do not give a deep analysis
    of those models. One of the common properties of practically all models is that they
    take into account statistical and systematic errors in  quadrature form and, in most part,
    give only a qualitative description of the behavior of the differential cross section
    in a wide momentum transfer region.


 However, there are two essentially  different ways of including statistical
 and systematic uncertainties in the fitting procedure, especially if we want to obtain
  a quantitative description of  experimental data.
 The first one, mostly used in connection with the differential cross sections  (for example \cite{BSW,Bourrely-14,Gotsman,Kohara}),
 takes into account  statistical and systematic errors in quadrature form:
    $\sigma_{i(tot)}^2 = \sigma_{i(stat)}^2+ \sigma_{i(syst)}^2$.
    In this case,  
     $\chi^2$  can be simply written as
\begin{eqnarray}
\chi^{2}=   \sum_{i=1}^{n} \frac{ ( \hat{E}_{i}  - F_{i}(\vec{a}) )^2  }
{\sigma_{i(tot)}^{2}} .
    \label{eq6}
 \end{eqnarray}

   The second approach accounts for the basic property of systematic uncertainties,
   i.e. the fact that these errors have the same sign and size in proportion to the effect
   in one set of experimental data and possibly  have a different sign and size in another set.
    To account for these properties, extra normalization coefficients
     $k_j = 1+ \sigma $
    for the measured data
     are introduced in the fit. For simplicity, this normalization is often transferred
     into the model parametrization
         $f_j = 1/k_j$,
       while it - in reality - accounts for the uncertainty of the normalization of  experimental data.
\cite{Sel-Or-18}.
     This method is often used by research collaborations to extract,
      for example, the parton distribution functions of nucleons
        \cite{exmp1-26,exmp1} and nuclei \cite{EPPS16}
       in high energy accelerator experiments, or in astroparticle physics  \cite{Koh15}.
       In this case, $\sigma_{i(tot)}^2 = \sigma_{i(stat)}^2$ and
   the systematic uncertainty are taken into account as an additional normalization coefficient,
    with inverse form $f=1/k$. \cite{Sel-Or-18}.
      Hence,  systematic errors can be represented as an additional
     normalization coefficient.  Then, only  statistical errors have to be taken into account
     in  calculations of $\chi^2$.
   \begin{eqnarray}
\chi^{2} =   \sum_{j=1}^{m}  [  \sum_{i=1}^{n} \frac{ (  \hat{E}_{ij}  - f_j F_{ij} )^2  }{ \sigma^{2}_{ij(st.)} }
 + \frac{(1-f_j)^2}{\sigma^{2}_{j}} ].
\end{eqnarray}
 It should be noted that in the minimization procedure used in these two methods,
       different sizes of experimental errors were assumed. In the first case,
        we account for experimental errors in the quadrature of  statistical and systematic errors
        and for experimental data with the normalization given by an experimental collaboration.
         In the second case, only statistical errors are considered as  experimental uncertainty.
          The systematic errors are accounted for as an additional normalization coefficient interpreted
          as a nuisance parameter applied to all experimental data of this separate data set.

   In the first case, the "quadrature form" of the experimental uncertainty gives a wide corridor
   in which different forms of the  theoretical amplitude can exist. In the second case,
   the "corridor of the possibility" is essentially narrow, and it restricts  different forms
   of theoretic amplitudes.

     To examine subtle effects in the behavior of  differential cross sections,
     it is needed to have the narrowest possible corridor  for testing a theoretical function.
     In  our paper \cite{Osc-13}, it  was shown that the new data of the TOTEM Collaboration at 13 TeV
     show the existence in the scattering amplitude of  the oscillation term, which can be determined
     by the hadron potential at large distances. In the analysis of experimental data
     of  both the sets of the TOTEM data the additional
     normalization was used. It size reaches sufficiently  large values. In this case, a very small
     $\chi^2_{dof}$ was obtained with taking into account only statistical errors and
     with  a small number of  free parameters in the scattering
     amplitude, which was obtained in our High Energy Generalized Structure (HEGS) model \cite{HEGS0,HEGS1}.
     However, the additional normalization coefficient reaches a sufficiently large value,
     about $13\%$. It can be in a large momentum transfer region but is very unusual for a small
     momentum transfer. However, both sets of experimental data (small and large region of $t$)
     overlap in some region and, hence,  affect  each other's normalization.
     It is to be noted, that the size of the normalization coefficient does not impact the size
     and properties of the oscillation term. We have examined  many different variants of our
     model (including large and unity normalization coefficient) , but the parameters of
     the oscillation term have small variations.

     In the present work, the analysis of both sets of the TOTEM data at 13 TeV
     is carried out with  additional normalization equal to unity and taking into account
     only  statistical errors in experimental data.
     Hence, the additional normalization coefficient in eq.(2)   are fixed by unity
      $f_j =1$. 
  In the work, the fitting procedure  uses the modern version of the program
    "FUMILIM" \cite{Sitnik1,Sitnik2}" of the old program
    "FUMILY" \cite{fum83} which calculates the covariant matrix and gives the corresponding errors of parameters and
    their correlations coefficients, and the errors of the final data.
    The analysis of the TOTEM data by three difference statistical  methods, including the calculations through the correlation matrix  of the systematic errors
     was made in \cite{GS-totan19}  

\begin{figure}
%
\begin{center} 
\includegraphics[width=.70\textwidth]{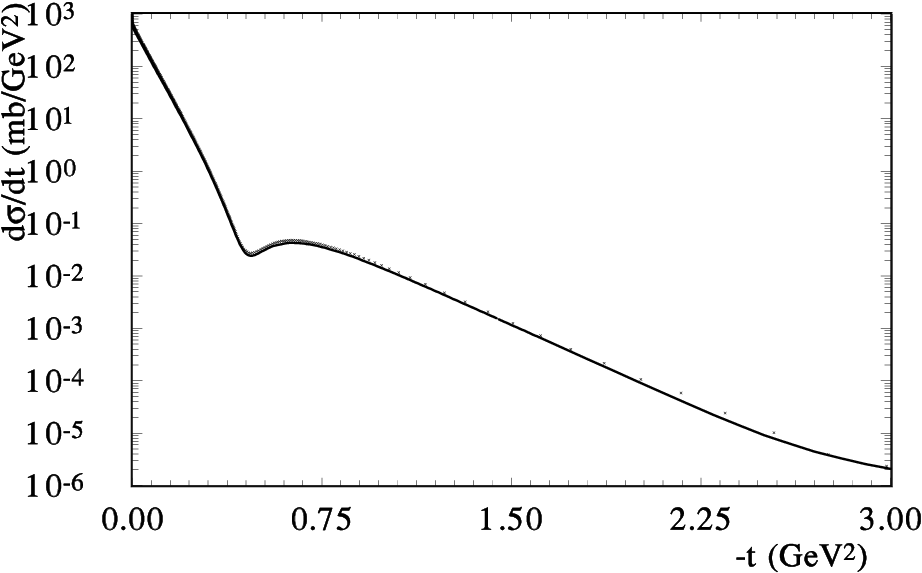}

 \vspace{1.cm}
 \includegraphics[width=0.45\textwidth]{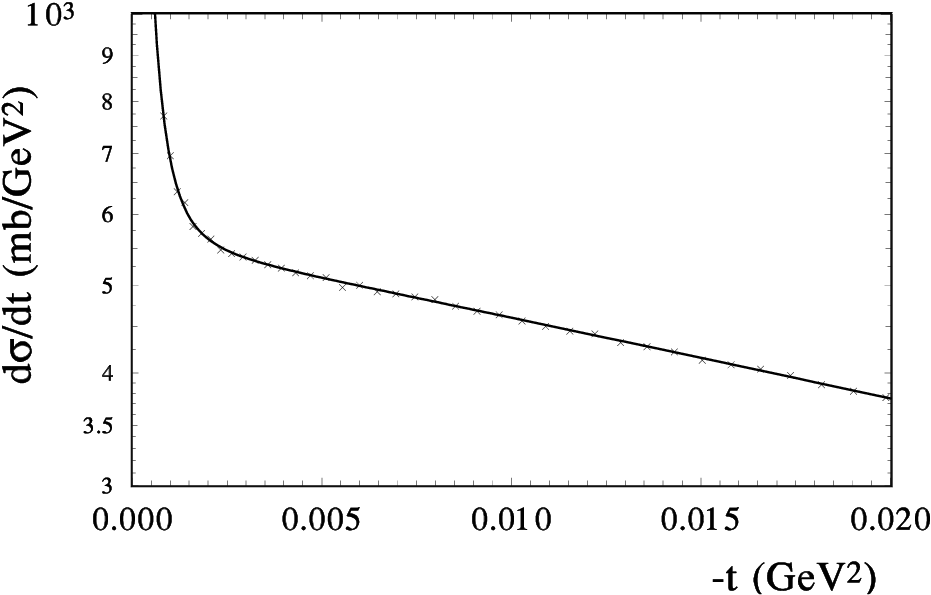}
 \includegraphics[width=0.45\textwidth]{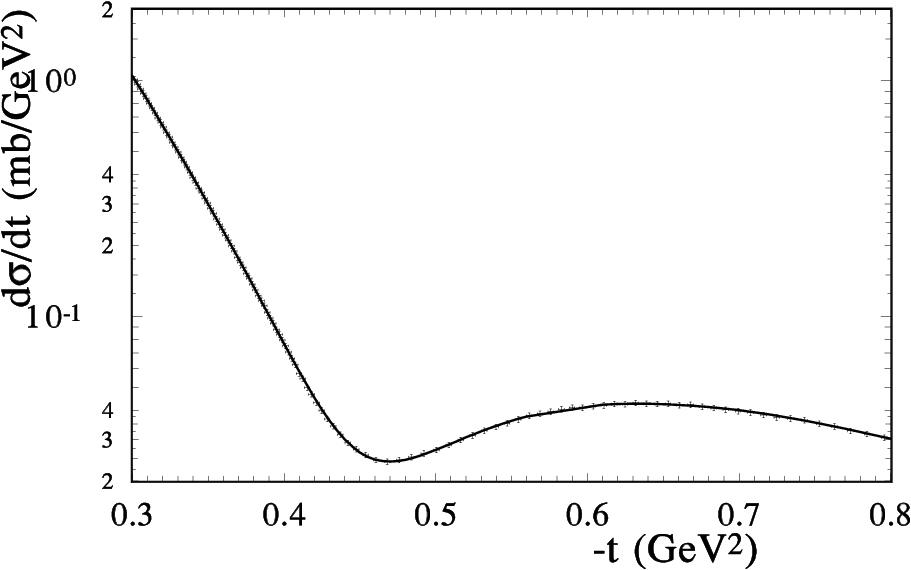}
\end{center}
\vspace{0.5cm}
\caption{The differential cross sections are calculated in the framework of the HEGS model
 with fixed additional normalization by $1.0$ and with additional term eq.(7),
  a) [top] the full region of $t$ and the data [1,2] 
   b) [bottom-left] the magnification of the region of the small momentum transfer of a);
  c) [bottom-right] the magnification  of the region of the diffraction minimum.
  }
\label{Fig_1a}
\end{figure}

\section{ Model description of two sets at 13 TeV with  additional normalization equal to unity}

   Differential cross sections measured experimentally
 are described by the squared scattering amplitude
\ba
d\sigma /dt &=& \pi \ (F^2_C (t)+ (1 + \rho^{2} (s,t)) \ Im F^2_N(s,t)
                                                            \nonumber \\
  & & \mp 2 (\rho (s,t) +\alpha \varphi )) \ F_C (t) Im F_N(s,t)).   \label{ds2}
\ea
where $F_{C} = \mp 2 \alpha G^{2}/|t|$ is the Coulomb amplitude;
$\alpha$ is the fine-structure constant,
$\varphi(s,t) $
 is the Coulomb hadron interference phase  between the electromagnetic and strong
 interactions (in our case, it is taken from \cite{selmp1,selmp2,selmp3}),
 and
$Re\ F_{N}(s,t)$ and $ Im\ F_{N}(s,t)$ are the real and
imaginary parts of the nuclear amplitude;
$\rho(s,t) = Re\ F(s,t) / Im\ F(s,t)$.
Just this formula is used to fit   experimental  data
determined by the Coulomb and hadron amplitudes and the Coulomb-hadron
phase to obtain the value of $\rho(s,t)$.

  As a basis, we take our high energy generalized structure (HEGS) model \cite{HEGS0,HEGS1} which quantitatively  describes, with only a few parameters, the   differential cross section of $pp$ and $p\bar{p}$
  from $\sqrt{s} =9 $ GeV up to $13$ TeV, includes the Coulomb-hadron interference region and the high-$|t|$ region  up to $|t|=15$ GeV$^2$
 and quantitatively well describes the energy dependence of the form of the diffraction minimum \cite{HEGS-min}.
   However, to avoid  possible problems
 connected with the low-energy region, we consider here only the asymptotic variant of the model.

   The total elastic amplitude in general receives five helicity  contributions, but at
   high energy it is enough to write it as $F(s,t) =
  F^{h}(s,t)+F^{\rm em}(s,t) e^{\varphi(s,t)} $\,, where
 $F^{h}(s,t) $ comes from the strong interactions and
 $F^{\rm em}(s,t) $ from the electromagnetic interactions.
 Note that all five spiral electromagnetic amplitudes are taken into account
 in the calculation of the differential cross sections.
    The Born term of the elastic hadron amplitude at large energy can be written as
    a sum of two pomeron and  odderon contributions,
    \begin{eqnarray}
 F_{\mathbb{P} }(s,t) & =& \hat{s}^{\epsilon_0}\left(C_{\mathbb{P}} F_1^2(t)  \
  \hat{s^{\alpha'} \ t} + C'_{\mathbb{P}} A^2(t) \ \hat s^{\alpha' t\over 4} \right) \; , \\
 F_{\mathbb{O} }(s,t) & =&  i \hat{s}^{\epsilon_0+{\alpha' t\over 4}} \left( C_{\mathbb{O} }
   + C'_{\mathbb{O}} \ t  \right) A^2(t).
 \end{eqnarray}
 All terms are supposed to have the same intercept  $\alpha_0=1+\epsilon_0 = 1.11$, and the pomeron
 slope is fixed at $\alpha'= 0.24$ GeV$^{-2}$.
  Many models  used
  the electromagnetic form factors of the hadron
  for the description of the scattering amplitude
 but, in most part, they changed their form to describe experimental data,
 as was made in  the famous  Bourrely-Soffer-Wu model \cite{BSW03}.
 The parameters of the obtained form-factor are determined by  fitting of the differential cross sections.
 The authors noted that the form factor is "parameterized like an electromagnetic form factor, as  two poles,
 and the slowly varying function reflects
 the approximate proportionality between the charge density
 and hadronic matter distribution inside a proton."

   In  paper \cite{Miettinen}, it was proposed that the  hadron form factor is proportional to the matter
   distribution. The matter distributions in the hadron are tightly connected with the energy momentum tensor
   \cite{Pagels}.  In \cite{Broniow}, it was noted that "the gravitational form factors are related to the matrix
   elements of the energy-momentum tensor in a hadronic state,
   thus providing the distribution of matter within the hadron".
   The recent picture of the hadron structure is determined
   by the general parton distributions (GPDs) \cite{Ji97,R97}
   which  include, as part, the parton distribution functions (PDFs). 

     In the HEGS model the form factors are determined by the general parton distributions of the hadron (GPDs) \cite{StrFF-PN14}.
      The first form factor, corresponding to the first momentum of GPDs is the standard electromagnetic
      form factor - $G(t)$. The second form factor is determined by the second momentum of GPDs -$A(t)$.
      The parameters and $t$-dependence of the GPDs are determined by the standard parton distribution
      functions and hence  experimental data on  deep inelastic scattering and by  experimental data
      for the electromagnetic form factors (see \cite{GPD-ST-PRD09}).

  The model takes into account  two hadron form factors $F_1(t)$ and $A(t)$, which correspond to  the charge and matter
  distributions \cite{GPD-PRD14}. Both form factors are calculated  as the first and second moments of  the same Generalized Parton Distributions (GPDs).  Hence,  additional fitting parameters are not required  for the description of the form factors. 

  The Born scattering amplitude has  four free parameters (the constants $C$) at high energy:
two for the two pomeron amplitudes  and two for the odderon.
The real part of the hadronic elastic scattering amplitude is determined
   through the complexification $\hat{s}=-i s$ to satisfy the dispersion relations.
   The oscillatory function was determined \cite{Osc-13}
\vspace{-0.1 cm}
\ba
 f_{osc}(t)=h_{osc} (i+\rho_{osc}) J_{1}(\tau))/\tau;  \ \tau = \pi \ (\phi_{0}-t)/t_{0},
\ea
where $J_{1}(\tau)$ is the Bessel function of the first order.
 This form has only a few additional fitting parameters and allows one to represent
 a wide range of  possible oscillation functions.

  After the fitting procedure we obtain  $\chi^2/n.d.f. =1.24$ (remember that we used only statistical errors).
 One should note that the last points of the second set above $-t=2.8$ GeV$^2$  show
 an essentially different slope, and we removed them. The total number of  experimental points
 of both sets equals $415$. If we remove the oscillatory function, then
  $\chi^2/n.d.f. =2.7$, so an increase is more than two times.  If we make a new fit without $f_{osc}$,
 then  $\chi^2/n.d.f. =2.4$ decreases but remains large.
  However, this result was obtained with a sufficiently large additional coefficient of the normalization
  $n=1/k=1.135$. It  can be for a large momentum transfer, but unusual for a small region of $t$.

  Now let us put the additional normalization coefficient to unity and continue to take into account
  in our fitting procedure only statistical errors. Of course, we obtain an enormously huge
   $\sum \chi^2$. The new fit changes the basic parameters of the Pomeron and Odderon Born terms
  but does not lead to a reasonable  size of  $\chi^2$.
  We find that the main part of    $\sum \chi^2$ comes from the region of a very small momentum transfer.
  It requires the introduction of a new term which can help to describe the CNI region of $t$.
   This kind of term can be taken in  different forms. In the present paper, we examined
   two different forms. One is the simple exponential form
  \ba
 F_{d}(t)=h_{d} (i+\rho_{d}) e^{-B_{d} |t|^{\kappa} \log{\hat{s} }},
 \label{fd-exp}
\ea
  and the other is the power form which has t-dependence similar to the squared Coulomb amplitude.
\ba
 F_{d}(t)=h_{d} (i+\rho_{d})/(1 + (r_{d} t)^2) \ G_{el}^2 .
  \label{fd-rb}
\ea
 where $ G_{el}^2 $ is the squared electromagnetic form factor of the proton.
  For simplicity, in a further fitting procedure  the constant $\rho_{osc}$  and the phase $\phi_{0}$
  of the oscillatory term are taken equal to zero. Hence, the  oscillatory term  depends only on two parameters -
  $h_{osc}$ and $t_{0}$ period of  oscillation. Also, to reduce the number of fitting parameters
  the correction to the main slope is taken in a simple form, we obtain the slope as
   \ba
 B(t)= \alpha^{'} \log{\hat{s}} ( 1- t e^{ B_{ad} t}).
 \label{slope-s}
\ea
    The Pomeron trajectory has threshold singularities, the lowest one
being due to the two-pion exchange required by the $t-$channel
unitarity \cite{Gribov-72}. This threshold singularity appears in different forms in various
models (see \cite{Jenk-72,Rev-LHC})  and is now recalculated by V. Khoze  \cite{Khoze-Sl00}.
    It can be shown that this term has a complicated structure and is determined by cancelation of two divergent terms
   with a small rest which we approximated by our small correction term to the main slope.
  This form leads to the standard form of the slope as $t \rightarrow 0$ and $t \rightarrow \infty$
  and practically does not effect the rapidly decreasing additional term $F_{d}(t)$.  

\begin{figure}
%
\begin{flushright} 
\includegraphics[width=.45\textwidth]{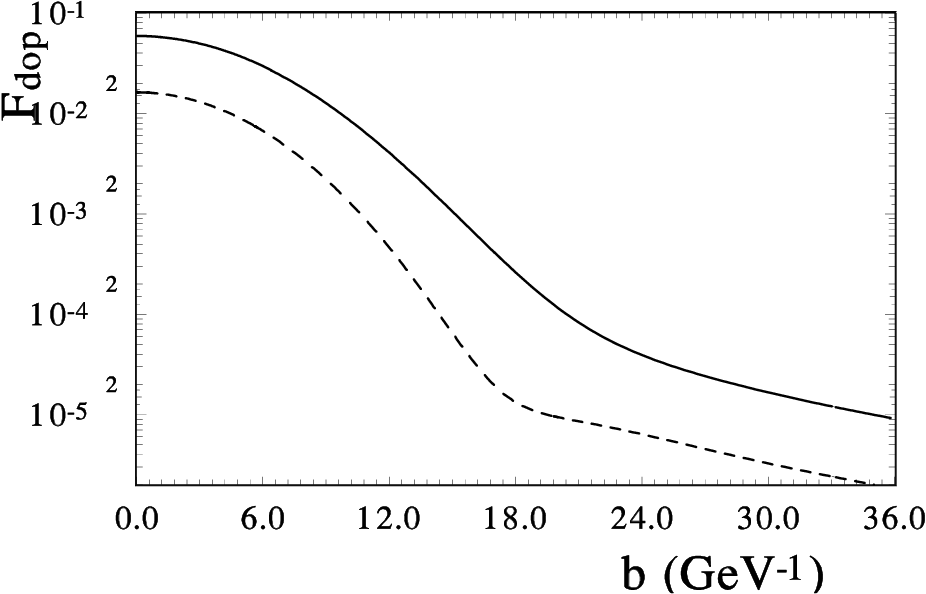}
\includegraphics[width=0.45\textwidth]{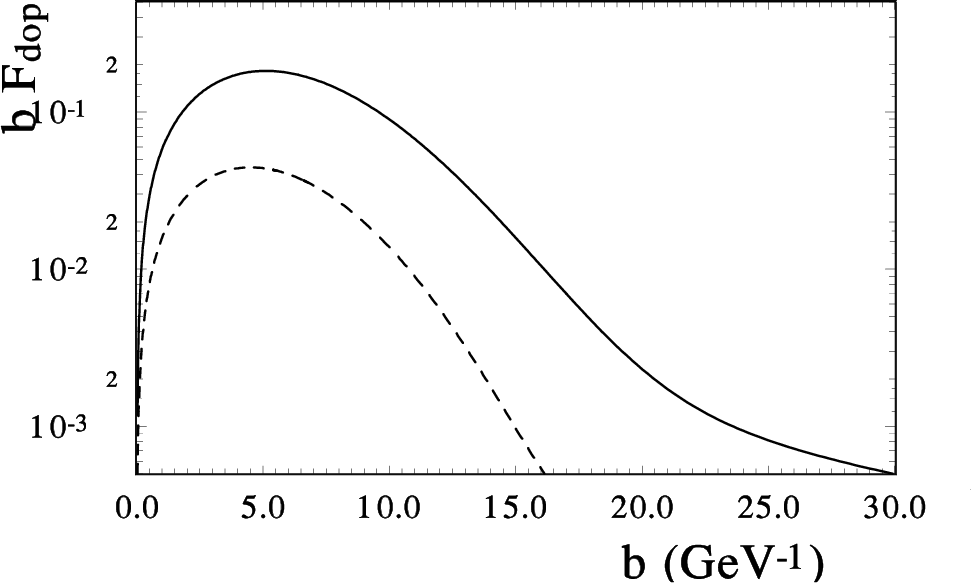}
\end{flushright}
\vspace{1.cm}
\caption{ The amplitude $F_{dop}(b)$ eq.(7) in the impact parameter representation
  a) [left] the real $F_{dop}(b)$- hard line and imaginary part $Im F_{dop}(b)$ -dashed line ;
  b) [right] overlapping function $b F_{dop}(b)$ (real part - hard line; imaginary part - dashed line) .
  }
\label{Fig_1}
\end{figure}


\begin{figure}
%
\begin{flushright}\vspace{-1.cm}
\includegraphics[width=.45\textwidth]{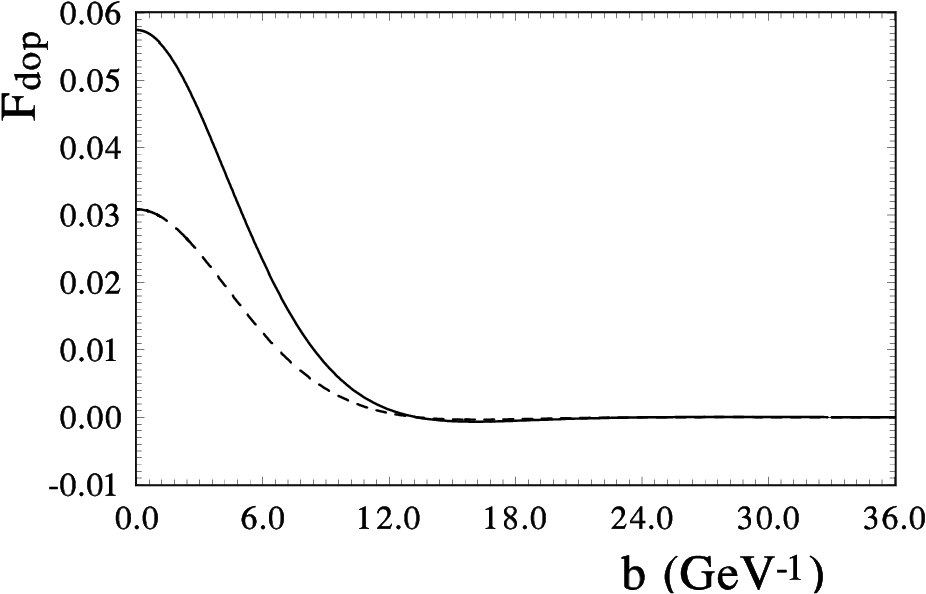}
\includegraphics[width=0.45\textwidth]{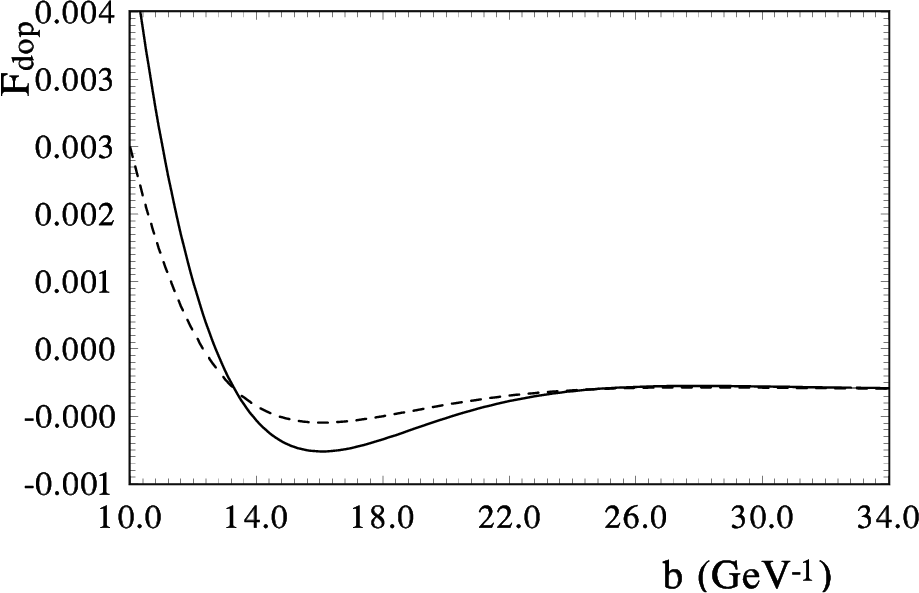}
\end{flushright}
\vspace{1.cm}
\caption{ The amplitude $F_{dop}(b)$ eq.(8) in the impact parameter representation
  a) [left] the real $F_{dop}(b)$- hard line and imaginary part $Im F_{dop}(b)$ -dashed line ;
  b) [right] the same at large impact parameters.
  }
\label{Fig_3}
\end{figure}

  The fit of both sets of the TOTEM data simultaneously with taking into account only statistical errors, with additional normalization equal to unity and with the additional term,  eq.(7), gives
  a very reasonable $\chi^2 = 551/425=1.29$. The results are present for the full region
  of $t$  in Fig.1a, and with zoom of the region of small $t$ in Fig.1b, and zoom of the region
  of the diffraction minimum in Fig.1c.

   The parameters of the additional term are well defined
   $ h_{d}=1.7 \pm 0.01;  \ \ \ \rho_{d}=-0.45 \pm 0.06; $ \\
   $ B_{d} = 0.616 \pm 0.026;  \ \ \  \kappa=1.119 \pm 0.024. $

   Using the second form of the additional term, eq.(8), we obtain
     practically the same picture with the same $\chi^2=549/425 =1.28$
    (  with the parameters of the additional term
 $ h_{d} = 1.067 \pm 0.044; \ \ \ \rho_{d} = -0.53 \pm 0.07 $ \\
 $ r_{d}= 7.62 \pm 0.34 $). \\
 To check up the impact of the form of the CNI  phase - $\varphi (t)$, we made our calculations
 with  the original Bethe phase $\varphi  = -( dLOG(Bsl/2.*t)+0.577)$ as well.
 We found that $\sum \chi^2$
 changes by less than $0.2\%$ and practically does not impact  the parameters $F_d (t)$.
  Hence, our model calculations show  two possibilities in the quantitative
  description of the two sets of the TOTEM data.
  One - takes into account an additional normalization coefficient, which
  has a minimum size of about $13\%$ ; the other - the introduction of a new anomalous term
  of the scattering amplitude, which has a very large slope and gives the main contributions
  to the Coulomb-nuclear interference region.

    Of course, there are some other ways to obtain good descriptions of the
    new experimental data of the TOTEM Collaboration. One is  to use  a model
    with an essentially increasing  number of the fitting parameters and many different
    parts of the scattering amplitude. Another is to use  a polynomial model
    with many free parameters.
    In both cases, the physical value of such a description is doubtful.


    Let us  examine the additional term in the impact parameter representation and
     use the Fourier transform
    \ba
 F_{dop}(b) \sim \int_{0}^{\infty} \ dq \ J_{0}(q b ) \ F_{dop}(q^2)  ,
 \label{rep-qb}
\ea
   The results for the additional term, in the form of eq.(7), are presented in Fig.2a.
    Figure 2b shows that the main contribution comes from the non-large impact parameters.
   The maximum of $b F_{dop}(b)$ is situated in the region of $r \sim 1 $fm, slightly above the electromagnetic radius of proton.
    Figure 3 shows  the impact parameter representation for the real and imaginary parts
   of $F_{dop}$, in the form of eq.(8).

\section{Other models}

  The  above results were obtained in the framework of one specific model.
   Let us see what  other models tell us. There are many different models with
   very different paradigms (for example, see reviews \cite{Rev-LHC,Pakanoni}) and we take only some
   of them as an example.
   One of the oldest models, which is based on the hadron structure
     \cite{Islam} 
 is enclosed by the quark-antiquark cloud.  The cloud becomes polarized
  because its antiquarks are drawn toward the baryonic shell.  In turn, a layer of polarization quarks appears.
     In pp near forward scattering, the two outer layers collide leading to a new scattering amplitude (positive).  In  $p\bar{p}$  near forward scattering, the outer polarization layer of the antiproton is of antiquarks and the polarization scattering amplitude is negative.Thus, polarization of the clouds incorporates a small crossing-odd amplitude into our diffraction amplitude.  
  It  says that  the main result is
 " The most striking feature of the preliminary $\sqrt{s} = 13 $ TeV TOTEM data is that there are no oscillations in
  $d\sigma/dt$ beyond the
initial dip-bump structure. It shows a smooth falloff for large $|t|$, exactly as predicted by our model."
 The model gives, as many others, only  a qualitative description of the differential cross section and does not feel
 the fine structure.

  Some other models, for example  \cite{Jnk-19}, developed the structure of the scattering
  amplitude, but in the analysis of the experimental data they do not include the specific properties
  of the hadron interaction at small momentum transfer -
 " Note that in this paper, we treat only the strong (nuclear) amplitude separated
  from Coulomb forces. The CNI effects modify the nuclear cross-section
by less than $1 \%$ for $|t| > 007$ GeV$^2$; thus, in the nuclear range, the CNI effects
can be ignored" 

The same specific bounds were taken  by one of the famous models  \cite{DL-19}.
  In a recent paper they noted: " This
paper applies it in its simplest form to small- $t$ data from $13.76$ GeV to $13$ TeV for total cross sections
and elastic scattering at small t, namely $|t| < 0.1$ GeV$^2$
by including in the amplitude the exchange of
the soft pomeron $|P$ of the reggeons $\rho, \omega, f2, a2$ and of two pomerons $IPIP$. The fit reveals no need[3]
for any odderon contribution at small t." 
As in many other papers in \cite{DL-19} they speak  only
about a good fit, which is shown in different Figures.
However, they do not speak about the sizes
of $\chi^2$, especially in different regions of momentum transfer.
It is interesting that in \cite{DL-19} they  note " ...but the slope for 7 TeV data lies between that for 8 and 13 TeV,
which is surely anomalous".
  Hence, all such models can not see some fine structure of hadron interactions, which
 is  discussed in our paper, but note some anomalous behavior of the slope.

  Some models include in the analysis the Coulomb-hadron interference region and
  note the importance of this region of $t$.
 However, in most part, they are interested in the deviation of the differential cross sections
 from the exponential form, which leads to some "break" in the region of $-t \sim 0.15$ GeV$^2$.
 For example, in  \cite{Koh-Bl18} they note -
"The left plot shows the non-exponential behavior of the differential cross
section for T8. .."
The figure is obtained subtracting from the best fit of the differential
cross section
 with a pure exponential
form $Re(f) = Aexp(Bt)$ and dividing the subtraction by this reference function. The
dashed lines show the normalization error band in $d\sigma/dt$, which is quite large.
The plot
in the RHS shows the ratio $(T2-R)/T2$
 which exhibits information of a non-exponential
behavior with advantages compared with the first plot, since  is cancelled, and
with it most of the normalization systematic error." 
 It is interesting that they show the importance of  systematic errors
 for a
  good description of the differential cross sections.
%


In a recent complicated work \cite{Pakanoni}, which based on the modified Barger-Philips  scattering amplitude \cite{Bar-Phil}, a qualitative description of the
experimental data at $\sqrt{s} = 7,8,13 $ TeV in the near-forward region up to $-t=0.2$ GeV$^2$  was obtained.
  One of the specific moment of the model is the use of the energy dependence of the
  hadron form factor.
  The obtained value of $\sigma_{tot}( \sqrt{s} = 13 TeV) = 113.66 $ mb
  in the first variant and $111.09$ in the fourth  variant of the model;
 the corresponding values of $\rho(t=0)$ equal $0.133$ and $0.134$.  

Another interesting model \cite{God} examines modern  experimental data of the TOTEM Collaboration.
They used   two-Pomeron eikonal approximation
with soft and hard Pomeron.
A specific feature of the model is that it uses the essentially nonlinear parametrization for the soft and hard pomerons of the Regge trajectory.
  With minimum fitting parameters obtained earlier ( via fitting to  elastic scattering data in the collision energy range
546 GeV $\le\sqrt{s}\le$ 7 TeV ) the model gives a good qualitative description of the new experimental data at 13 TeV.
After refitting parameters, it gives for the data at 13 Tev $\sum \chi^2=980$.
 The result is very interesting from our viewpoint. We present their Fig. 2b in our Fig.4.
 The difference between the model results and the experimental data at small
 momentum transfer is remarkable. Very likely that it shows the necessity of
additional normalization of the experimental data or the existence of some anomaly in  $t$ dependence of the differential cross sections.  
=

\begin{figure}
\begin{center}
\includegraphics[width=.7\textwidth]{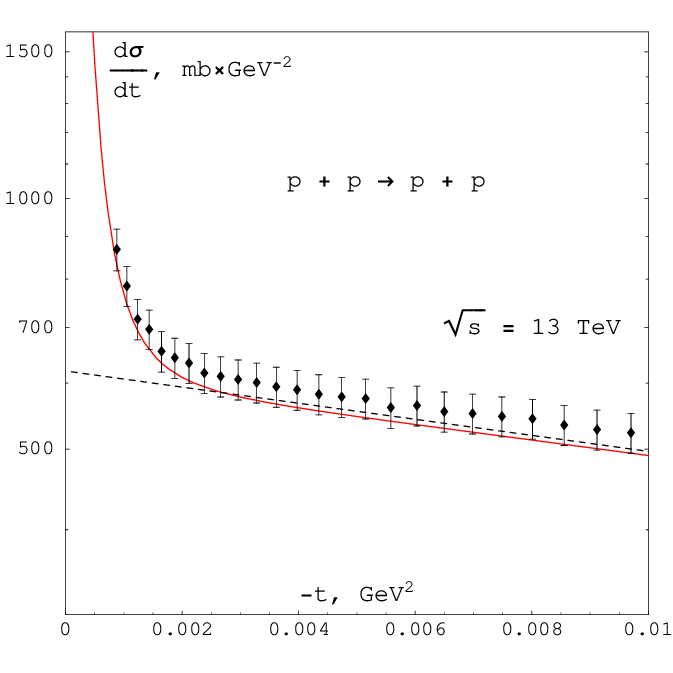}
\end{center}
\caption{from paper [39] 
 "Figure 2: The predictions of the model [6] in the case  $HP(0) - 1 = 0.32$ versus the TOTEM
 data at $\sqrt{s}=13$ TeV [7]. The dashed line corresponds to the approximation $C(s,t) = 0$" }
\label{Fig_6a}
\end{figure}

\section{The fit of the differential cross section in the small momentum transfer region}

   Above, the examination of the new TOTEM experimental data at $\sqrt{s}=13$ TeV carried out in a wide
   momentum transfer shows the existence of some anomaly in the behavior of the differential cross sections
   at a very small momentum transfer.  Of course, it has some dependence on  the model structure.
   We cannot exclude a possibility of discovering  a more complicated model that explains new features
   of  hadron interactions at large distances.
   Hence, it is important also to see the phenomena of   the new effect only in the small momentum transfer region and in the framework of the simplest form of the scattering amplitude .
   Now let us limit our examination to a small region of momentum transfer (up to $-t=0.069$)
   which includes 79 experimental data of the first set of the TOTEM Collaboration \cite{T66}.
    This region was examined by the TOTEM Collaboration 
   and some other groups of researchers (for example \cite{CS-Diff,Protvino19}).
   Unlike other groups, we will  take into account only statistical errors and  the
   additional normalization $k=1$. The new data of the TOTEM Collaboration have  very small statistical
   errors, especially in the low momentum region. Hence, our fitting procedure will give the non-small
     $\sum \chi^2_{i}$; however, it imposes hard restrictions  on  different representations of the scattering amplitude.
    Firstly, let us examine the Born scattering amplitude using the standard eikonal representation,
    as was made in our model analysis of the whole region of the momentum transfer of experimental data.

       Let us take the hadronic Born scattering amplitude in the simple exponential form. Of course,
       after eikonalization such an amplitude is added  the  standard electromagnetic amplitude,
       as we made in the model analysis,
\begin{eqnarray}
 F_{\mathbb{P} }(s,t) = h/(2 \ 0.389 \pi ) (i+\rho)  Exp( B/2 t)
 \end{eqnarray}
   As was made in a recent work \cite{Protvino19}, we will made the fit
   in  different regions of $t$.
   Our results are given in  Table 1.
Table 1 and the following tables show the sizes of $\sum \chi^2$ and
    the integrated probability - $p-value$
   (the area under $\chi^2$ probability density function (pdf) to the right of the minimum $\chi^2$ value
   (see, for example, \cite{Xudson}). 
    The maximum width of the examined region leads to the non-small
   $\sum \chi^2_{i}$. It is shown that a simple exponential form is not sufficient for  our analysis.
   Of course, when we come to a small region of $t$, the description is improving more and more.
   It is to be noted that the size of the slope has small variation with decreasing  $t$.
   In our analysis, the slope size is somewhat   less than was determined by the TOTEM group \cite{T66} and
   by the Protvino group \cite{Protvino19}. This may be the  result of  the slope determined by the Born
   scattering amplitude that is further changed by the eikonalization procedure.
   However, we are interested in the possibility of the contribution of an additional
   rapidly decreasing term of the scattering amplitude.

   Let us add an additional term in the form
  \begin{eqnarray}
 F_{ad }(s,t) =  i \ h_{d}/(2 \ 0.389 \pi ) \ e^{ D_{d} t}
 \end{eqnarray}
This form correspond to eq.(7)
  but with some simplification as the narrow region of momentum transfer
   requires  minimum fitting parameters. 
   As a result, two additional parameters appear in the fitting procedure. We reduce the real part of the additional term as the increase in the number of the fitting parameters leads
   to large uncertainty in the results.
     It should be noted that if we add two additional parameters in the main Born amplitude
     as additional slopes - $B_{1} \ \sqrt(-t)$ and $B_{2} t^{2} $, this will not practically
      change the picture. The same result was obtained by the TOTEM Collaboration, too.

  The results of our new fitting are presented in Table 2.
  The $\chi^2$ decreases is essential, especial for the completely examined $t$ region.
  The constant of the additional term is determined sufficiently well.
  The slope of the additional term is large and also is determined with small errors.

\begin{figure}
\begin{center}
\includegraphics[width=.8\textwidth]{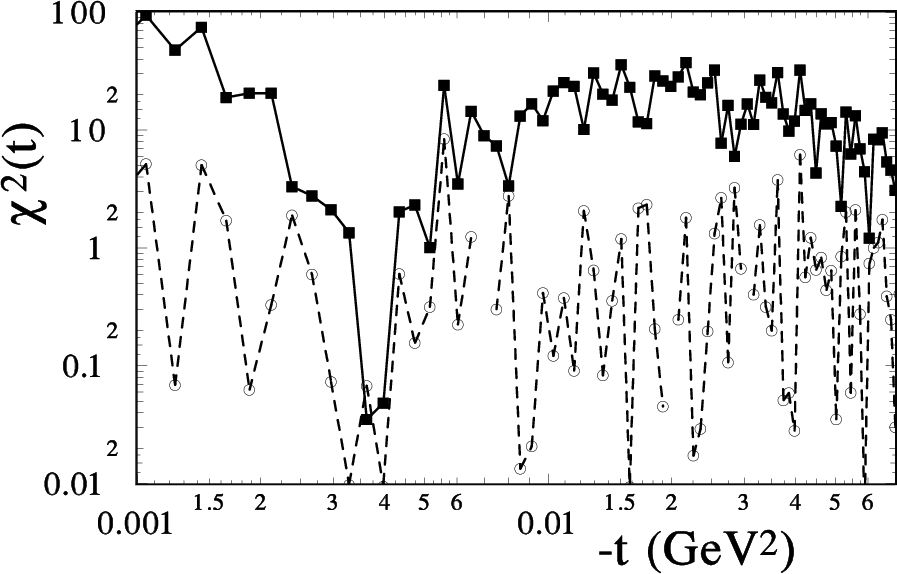}
\end{center}
\vspace{1.cm}
\caption{ The $\chi^{2}(t) $  in the case of taking into account the additional fast decreasing term
 (dashed line) and in the case of the absence of such a term (hard line).
  }
\label{Fig_5}
\end{figure}


\begin{figure}
\begin{center}
\includegraphics[width=.45\textwidth]{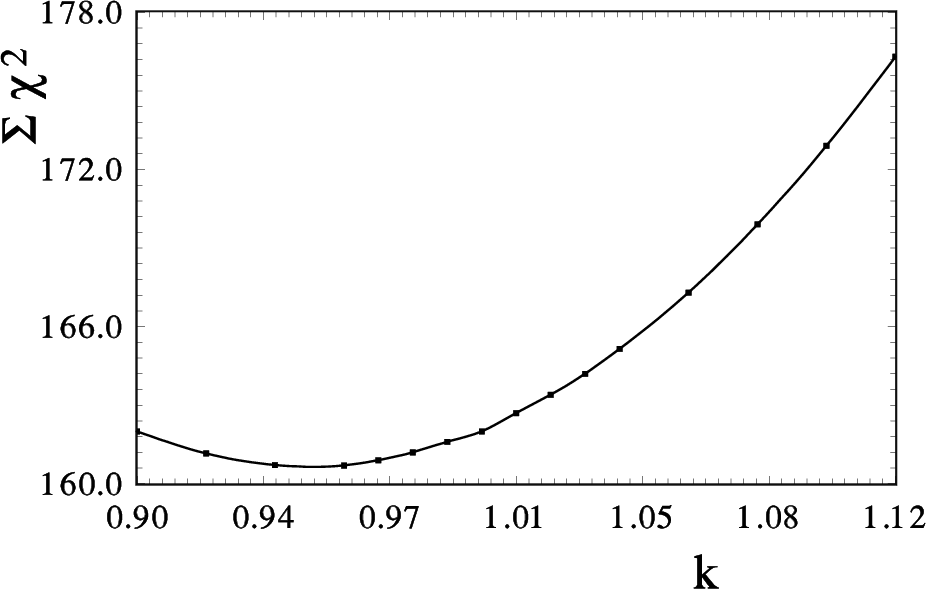}
  \includegraphics[width=0.45\textwidth]{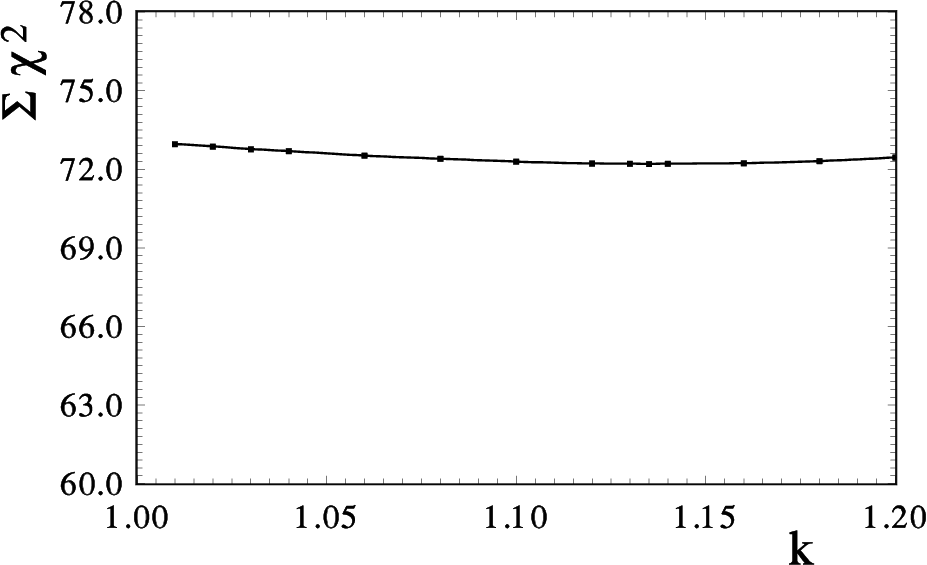}
\end{center}
\vspace{0.5cm}
\caption{ The dependence of  $\chi^2$  on the additional normalization  coefficient
  of experimental data -
  a) [top] in the case of the simple exponential form  of the scattering amplitude;
  b) [bottom] the same but with the additional fast decreasing term.
 }
\label{Fig_6}
\end{figure}
\vspace{0.5cm}


The $\chi^{2}(t) $  are shown in Fig.5   in the case of taking into account the additional fast decreasing term
 (dashed line) and in the case of the absence of such a term (hard line).
 We can see that the largest difference comes from a very small region of momentum transfer.

 Let us include  additional normalization (representing systematical errors)  in our fitting
 procedure. The dependence of  $\sum_{i}^{N} \chi^2$  on the additional coefficient
of  normalization of  experimental data is shown in Fig. 6.
 The case of the simple exponential for the scattering amplitude is presented at the top of Figure 4;
  and  with the additional fast decreasing term at the bottom. 
One  can see that  $\chi^2$ in the first case essentially depends on the normalization coefficient
 and has a sufficiently large value (remember that we used only statistical errors).
  Note that $k=1/n$ is the coefficient by which we multiply  our theoretical function
 to compare with  experimental data in our fitting procedure. The minimum is reached when
 the additional normalization equals $13.5\%$. This corresponds to the additional normalization
 which was used in our HEGS model calculations without the additional fast decreasing term.
 Contrary, in the case with the additional term the dependence on normalization is weak
 and the size of $\chi^2$ has a reasonable value in a wide region of normalization.

\begin{table}
\tbl{The fit of $d\sigma/dt$ by the Born hard scattering amplitude in  one \hspace{0.2cm}
 \hspace{1.3cm} exponential form   using the eikonal representation.}
{\begin{tabular}{@{}|c|c|c|c|c||c|c|c@{}} \toprule
 N&$-t_{max}(GeV^2)$&$\sum \chi^2_{i}$& $\chi_{dof}$ & pdf &h (GeV$^{-1})$   & $B (GeV^{-2} )$   \\ \hline
 79 & 0.0699 & 162.1& 2.19 & 2.5E-8   &$110.2 \pm 0.3$ & $16.0 \pm 0.03$ \\
 70 & 0.0559 & 86.5 & 1.29 & 0.055&$110.5 \pm 0.4$ & $ 16.4 \pm 0.04$ \\
 65 & 0.0488 & 66.3 & 1.07 & 0.331&$110.6 \pm 0.6$ & $16.3 \pm 0.05$ \\
 60 & 0.0422 & 55.3 & 0.97 & 0.539&$110.6 \pm 1.4$ & $16.4 \pm 0.06$ \\
 55 & 0.0361 & 49.7 & 0.96 & 0.565&$110.6 \pm 1.7 $& $ 16.4 \pm 0.08 $ \\
 50 & 0.0305 & 47.8 & 1.02 & 0.440&$110.6\pm 1.4 $ & $ 16.4 \pm 0.1 $ \\
 40 & 0.0207 & 34.2 & 0.92 & 0.601&$110.7 \pm 0.6$ & $16.6 \pm 0.2$ \\   \hline 
\end{tabular} \label{Table-1} }
\end{table}
\vspace{.5cm}


\begin{table}
\tbl{The fit of $d\sigma/dt$ by the Born hard scattering amplitude in  two \hspace{0.5cm}
\hspace{1.3cm} exponential form   using the eikonal representation.}
{\begin{tabular}{@{}|c|c|c|c|c||c|c|@{}} \toprule
 N&$-t_{max}(GeV^2)$&$\sum \chi^2_{i}$& $\chi_{dof}$ & pdf & h (GeV$^{-1})$   & $D_{d} (GeV^{-2} )$   \\ \hline
 79 & 0.0699 &  74  & 1.00&  0.478& $3.26 \pm 0.3$ & $41.2 \pm 1.9$ \\
 70 & 0.0559 & 62.2 & 0.93&  0.575& $2.94 \pm 0.4$ & $39.2 \pm 4.1$ \\
 65 & 0.0488 & 56.8 & 0.94&  0.593& $2.26 \pm 0.6$ & $31.6 \pm 5.8$ \\
 60 & 0.0422 & 53.0 & 0.96&  0.550& $1.51 \pm 1.4$ & $25.3 \pm 7.7$ \\
 55 & 0.0361 & 49.0 & 0.98&  0.513& $1.56 \pm 1.7 $ & $ 25.5 \pm 11.4 $ \\
 50 & 0.0305 & 44.4 & 0.98&  0.497& $1.93 \pm 1.4 $ & $ 29.7 \pm 13.7 $ \\
 40 & 0.0207 & 33.0 & 0.94&  0.565& $2.4 \pm 0.6$ & $29.7_{fixd}$ \\  \hline 
\end{tabular} \label{Table-2} }
\end{table}

%
\vspace{.5cm}

\begin{table}
 \tbl{The comparison of $\sum \chi^2_{i}$ from the fit of $d\sigma/dt$  by the hard scattering \hspace{0.2cm}
\hspace{1.3cm}amplitude in the exponential and two exponential forms. }
{\begin{tabular}{@{}|c|c|c|c||c|c|c|c|@{}} \toprule
 N&$-t_{max}(GeV^2)$&$\sum \chi^2_{i}$ (Exp) &pdf &  $\sum \chi^2_{i}$ (Exp+fd)&pdf & $h_{d}$    \\ \hline
 79 & 0.0699 & 67.61 & 0.686 & 62.87 &  0.818 &$ 1.95\pm 0.35$ \\
 70 & 0.0559 & 61.52 & 0.600 & 59.24 &  0.678 &$ 2.14\pm 0.58$ \\
 65 & 0.0488 & 57.14 & 0.628 & 55.32 &  0.647 &$ 2.25\pm 0.73$ \\
 60 & 0.0422 & 54.51 & 0.490 & 52.90 &  0.550 &$ 2.36\pm 0.95$ \\
 55 & 0.0361 & 50.26 & 0.460 & 48.39 &  0.498 &$ 2.03\pm 0.82$ \\
 50 & 0.0305 & 45.22 & 0.463 & 41.67 &  0.613 &$ 1.35\pm 0.54$ \\
 45 & 0.0254 & 38.03 & 0.569 & 34.58 &  0.712 &$ 2.33\pm 1.35$ \\
 40 & 0.0207 & 35.02 & 0.467 & 32.45 &  0.592 &$ 1.95\pm 1.35$ \\  \hline 
\end{tabular}  \label{Table-3}  }
\end{table}
\vspace{.5cm}

  Now let us carry out analysis without  eikonalization.
  In this case, the additional term will be represented in the power form (like a square of Coulomb amplitude)
    \begin{eqnarray}
 F_{ad }(s,t) = \alpha_{el}^2 \ h_{d}/(2 \  0.389 \pi ) /[ \epsilon + t^2]
 \end{eqnarray}
 where  $\alpha_{el}=1/137$ is the electromagnetic fine structure constant and $\epsilon$ is free parameter
 order   $\alpha_{el}^2$.
  This form corresponds to eq.(8) but in the form more close
  to the screening  Coulomb amplitude, as
   it is remarkably close to the form of the Coulomb amplitude but without the divergence at $t \rightarrow \infty$. 
  The comparison of  $\chi^2$ for a simple exponential term
 and with the added fast decreasing term are presented in Table 3. The difference is not large;
  however, it is about $10\%$
 for every examined region of $t$. The constant $h_{d}$ is also well determined.

\section{Conclusion}

     Using only statistical errors and fixing  additional normalization
      of differential cross sections equal to unity,
     we have limited the possible forms of the theoretical representation of the scattering
     amplitude.
     The phenomenological model - HEGSh model was used
     for examining the whole region of the momentum transfer
     of two sets of experimental  data obtained by the TOTEM Collaboration at $13$ TeV.
      The simple exponential form of the scattering amplitude was used to examine only
     a small region of  momentum transfer.
     In both cases, an additional fast decreasing term of
     the scattering amplitude was required for a quantitative description of the
     new experimental data.
      The large slope of this term can be connected with a large radius of the hadronic
      interaction and, hence, can be determined by the interaction potential at large distances.
      It can be some part of the hadronic potential responsible for the oscillation behavior
      of the elastic scattering amplitude \cite{Osc-13}.

      The discovery of  such anomaly in the behavior of the differential
      cross section at very small momentum transfer 
      in  LHC experiments will give us  important information about
      the behavior of the hadron interaction potential at large distances.
      It may be tightly connected with the problem of  confinement.
      We have shown the existence of such anomaly
      at the statistical  level and  that some other models also
      revealed such unusual behavior of the scattering amplitude.
      Very likely,  such effects exist also in  experimental data
      at essentially smaller energies \cite{osc-conf}.
      However, the results of the TOTEM Collaboration have a unique
      unprecedentedly small statistical error and reach minimally  small
      angles of  scattering with the largest number of  experimental poits
      in this small region of the momentum transfer.
     The new effects can impact  the determination of
     the sizes of the total cross sections, the ratio of
    the elastic to the total cross sections and the size of the
    $\rho(s,t)$ - the ratio of the real to imaginary part of the elastic scattering
    amplitude.

    Now the results for the total cross sections and $\rho(t=0)$ can be compared
    for the case with additional coefficient normalization $k$ and  the cases with
    an additional fast decreasing term and $k=1$.
    The results are presented in Table IV. The different variants with a large coefficient of the normalization
    give practically the same value, which is less than  the total cross sections  extracted by the TOTEM Collaboration
    - $\sigma_{tot (TOTEM)} =110.6 \pm 3.4$  mb in the analysis of only  small momentum transfer region
    \cite{T-st}.
  Small errors of $\rho(t=0)$ and $\sigma_{tot}$ are the result of our  simultaneous fitting to
    both sets in a wide region of the  momentum transfer and with using only statistical errors.
    The size of $\rho(t=0)$ obtained in the model calculations essentially exceed
    the size of $\rho(t=0)=0.1 \pm 0.01$ extracted by the TOTEM Collaboration \cite{T-rho}.
    On the contrary, the variants with an additional fast decreasing term
    in  different forms  give a
    large value of $\sigma_{tot}(\sqrt(s)=13$ TeV which exceeds the  $\sigma_{tot(TOTEM)}$,
    and $\rho(t=0)$ practically coincides with the predictions of the COMPETE Collaboration
    \cite{COMPETE}.
 Now many different groups using the TOTEM Collaboration data obtained
     different results for $\rho(t=0)$ and $\sigma_{tot}$. When they used only small -t data,
    the results were not far from the TOTEM   data for  $\sigma_{tot}$ but essentially differed
    for $\rho(t=0)$ (see, for example, \cite{Protvino19}).  

\begin{table}
 \tbl{A comparison of $\sigma_{tot} (\sqrt(s)=13 \ TeV)$ and 
$\rho(t=0,\sqrt(s)=13 \ TeV)$ obtained in the different variants
of the model calculations. }
{\begin{tabular}{@{}|c|c|c|c|c|c|@{}}      \toprule              
 n     & model          & $\sum \chi^2_{i}$/N & $10^4  \times$ pdf  & $\sigma_{tot}\pm err$ (mb) &  $ \rho(t=0)\pm err$     \\   \hline
 1.135 &                & 525/415 &$0.8 $& $106.1 \pm 0.2$     & $ 0.146 \pm 0.004 $     \\
 1.135 &$f_{d}=0$       & 515/425 &$6.5 $& $106.2 \pm 0.2$     & $ 0.142 \pm 0.004 $     \\
 1.135 &                & 527/425 &$2.3 $& $106.2 \pm 0.2$     & $ 0.148 \pm 0.004 $     \\ \hline
 1.    &$f_{d}(r_{d})$  & 539/425 &$0.3 $& $113.2 \pm 0.106$   & $ 0.109 \pm 0.004 $     \\
 1.    &$f_{d}(r_{d})$  & 549/425 &$0.1 $& $113.1 \pm 0.106$   & $ 0.113 \pm 0.004 $       \\
 1.    &$f_{d}(Exp)$    & 550/425 &$0.1 $& $112.6 \pm 0.107$   & $ 0.115 \pm 0.004 $       \\ \hline
\end{tabular} \label{Table-4} }
\end{table}
%

In \cite{1812.04732}  the TOTEM Collaboration notes that  the two combined TOTEM results yield
   $\sigma_{tot} = 110.5 \pm 2.4$ mb. It means that  our model calculations (see Table 4) differ by two
   $\sigma_{errTOTEM}$ for the case with large additional normalization and by one $\sigma_{errTOTEM}$
   for the case where the additional normalization is fixed by unity. Note that in the model calculations
   only statistical errors were used.

    Of course, we can not exclude the case that the real experimental normalization
    reaches essentially larger values than taken into account by the TOTEM
    Collaboration. However, for a small momentum transfer it is a very unlikely 
    case, as practically in all existing experiments on  measurement of the differential
    cross sections at the small momentum transfer  systematic errors
    do not exceed a few percent.


\vspace{0.5cm}

{\bf Acknowledgements}
 {\it The authors would like to thank Jean-Rene Cudell
	  for fruitful   discussion of some questions   considered in the paper.}

\end{document}